# Solenoidal Coils Made from Monofilamentary and Multifilamentary MgB$_2$ strands

M.D.Sumption[1], M. Bhatia[1], F. Buta[1], S. Bohnenstiehl[1], M. Tomsic[2], M. Rindfleisch[2], J. Yue[2], J. Phillips[2], S. Kawabata[1,3], and E.W. Collings[1]

[1]LASM, Materials Science and Engineering Department,
OSU, Columbus, OH 43210, USA
[2]Hyper Tech Research, Inc. Columbus, OH 43210, USA
[3]Kagoshima University, Kagoshima, Japan

**Abstract**

Three solenoids have been wound and with MgB$_2$ strand and tested for transport properties. One of the coils was wound with Cu-sheathed monofilamentary strand and the other two with a seven filament strand with Nb-reaction barriers, Cu stabilization, and an outer monel sheath. The wires were first S-glass insulated, then wound onto an OFHC Cu former. The coils were then heat treated at 675°C/30 min (monofilamentary strand) and 700°C/20 min (multifilamentary strand). Smaller (1 m) segments of representative strand were also wound into barrel-form samples and HT along with the coils. After HT the coils were epoxy impregnated. Transport $J_c$ measurements were performed at various taps along the coil lengths. Measurements were made initially in liquid helium, and then as a function of temperature up to 30 K. Homogeneity of response along the coils was investigated and a comparison to the short sample results was made. Each coil contained more than 100 m of 0.84-1.01 mm OD strand. One of the 7 strand coils reached 222 A at 4.2 K, self field, with a $J_c$ of 300 kA/cm$^2$ in the SC and a winding pack $J_e$ of 23 kA/cm$^2$. At 20 K these values were 175 kA/cm$^2$ and 13.4 kA/cm$^2$. Magnet bore fields of 1.5 T and 0.87 T were achieved at 4.2 K and 20 K, respectively. The other multifilamentary coil gave similar results.

**Introduction**

Many groups now fabricate $MgB_2$ wires [1-13], powder-in-tube-processed (PIT) strands being favored for long length applications. In some cases the $MgB_2$ is in direct contact with Cu, in other cases Fe, Nb, or other chemical barriers are used. Typical outer sheath materials are stainless steel, monel, Cu-Ni, and Cu. There are two main variants of PIT $MgB_2$ fabrication: ex-situ [1-4], and in-situ [6,10-13]. Numerous efforts to develop $MgB_2$ strand are ongoing, and significant progress is being made in improving its basic properties such as transport $J_c$, upper critical field, and irreversibility field. $MgB_2$ has the potential to fill an important niche as an inexpensive, lightweight conductor usable for coils operating at temperatures of up to about 30K.

An important application of $MgB_2$ is for the windings of MRI magnets. Here two classes of magnet are of interest -- 0.5 T systems (competing with resistive and permanent magnet systems) and higher field (2-4 T) systems. $MgB_2$ conductors are expected to operate at temperatures of 20-30 K, and be available at a lower cost per meter as well as a lower specific ($/kAm) cost than the Bi- or Y- based high $T_c$ conductors currently being considered. Persistent joints, while only now in development, are not expected to be as difficult as those for Bi- or YBCO-based conductors. Finally, $MgB_2$ conductors will offer a good balance between stability and protect-ability for coil applications.

**Background**

Soltanian, Dou, et al., [14] have made small solenoids from monofilamentary, 1 mm OD, in-situ based $MgB_2$/Cu wires. The strand, of total length 3 m, reached a transport critical current, $I_c$, of 72 A (1.3 x $10^5$ A/cm$^2$) at self field and 4 K. Bhatia et al. have measured 1 m lengths of $MgB_2$/Cu and $MgB_2$/Fe/monel strands wound into helical samples (ITER barrels) [15], reaching $10^4$ and $10^5$ A/cm$^2$ at 4.2 K and 4 T, respectively. A somewhat larger solenoid coil was made by Tanaka, Togano, et al. (Hitachi) [16] using a monocore, ex-situ powder-based, $MgB_2$/Ni tape. The coil required 10 m of wire (80 turns), was wax impregnated, and reached an $I_c$ of 105 A at 4.2 K. Machi and Murakami [17] fabricated a solenoidal coil using 3.5 m of monofilamentary, 0.5 mm OD, in-situ based $MgB_2$/Cu which gave 76 A at 4.2 K in self field (4.4 x $10^5$ A/cm$^2$). Fang, Salama, et al [18] fabricated a squat solenoid with a total wire length of 4 m wound from 1 mm square, monofilamentary, in-situ powder-based $MgB_2$/Fe. The coil had an $I_c$ of 185 A at 4.2 K and 1 T. A solenoid fabricated by Hyper Tech Research (reported by Hascicek et al [19]) required 20 m of wire. The round strand, 1 mm OD, of in-situ powder-based $MgB_2$/Cu achieved 278 A at 4.2 K and self field. Its 170 turns were sol-gel insulated. Serquis, Civale, et al. [20] fabricated a solenoid using 25 m of 1 mm OD wire of ex-situ $MgB_2$ powder in a stainless steel sheath; it achieved an $I_c$ of 350 A in self field at 4.2 K. Musenich et al [21] reported on pancake coils wound from tape which carried 347 A at 4.2 K in the coil and generated 0.75 T.

Previously, we have reported results on racetrack coils, where 120 A was achieved at 4.2 K in a coil with 80 turns of monofilamentary wire [22, 23]. In this work, we describe the fabrication and testing of three solenoids, each wound with 100+ m of strand. The strand type used for these coils includes both monofilamentary Cu-based wire and a seven filament MF conductor with Cu stabilization. Below we detail the strand manufacture, coil winding, HT, epoxy impregnation, testing, and analysis For two of these coils, the $I_c$ reached 220+ A, at 4.2 K and 126 A at 20 K, representing 100% of the short sample value. Fields of 1.5-1.57 T at 4.2 K and 0.87-0.97 T at 20 K were achieved.

**Strand Fabrication**

The continuous tube forming/filling (CTFF) process was used to produce $MgB_2$/Cu composite strands. This process, as developed at Hyper Tech Research (HTR), begins with the dispensing of powder onto a metal strip as it is being continuously formed into a tube. The starting 99.9% Mg powders were -325 mesh (but had an approximate top size of 20 µm), and the 99.9% B powders were amorphous, at a typical size of 1–2 µm. The powders were V-mixed and then run in a planetary mill. Two powder types were made: stoichiometric binary powder and Mg rich powders. For the Mg rich powders, Mg was added to the initial Mg and B mixture forming the ratio **$Mg_{1.1}B_2$**. During the forming/filling process, either Cu or Nb strip was used to encapsulate the powders. After exiting the mill at a diameter of 5.9 mm the filled overlap-closed tube was inserted into a full hard 101 Cu tube. For the Cu-monofilamentary strand this was then drawn to final size. For multifilament construction, this Nb-plus-Cu clad monofilament was drawn to the proper size and then restack as a 6 around 1 (Cu) in a monel outer can and drawn to final size. Strand specifications are given in Table 1. For further details on these multifilamentary strands, see [24].

**Coil Winding, Heat Treatment, Epoxy Impregnation**

The former was solenoidal and made from OFHC Cu. The strands for all three coils were insulated with S-glass insulation. The coils had from 364 to 538 turns of strand, see Table 2. Cu-1 was HT for 675°C/30 min, while NbCu-7A and B were HT for 700°C/20 min. The ramp up time was 2.5 h and the ramp-down time was approximately 5-6 h, and all HT were performed under flowing Ar. Coils Cu-1 and NbCu-7A were vacuum impregnated with mixed Stycast 1266 epoxy heated to 40°C. NbCu-7B was merely dipped into degassed epoxy (40°C). After removal from the epoxy bath the coil curing was performed in air (at room temperature). Total curing time was estimated at 6-12 h.

**Coil Measurement and Results**

Transport properties of the coils were measured in a LHe cryostat (Figure 1 shows Cu-1 mounted and ready for insertion). The 4.2 K measurements were performed in liquid He, while higher temperature measurements were made as the coil warmed up. Two Cernox temperature sensors were mounted on the coil, one on the top and one on the bottom. The temperature difference across the coil was never greater than 0.3 K. Voltage taps were placed an various places along the winding. The typical distance between successive taps was about 14-20 m. The field was measured with a cryogenic hall probe and a Bell gaussmeter calibrated to achieve a 2% or better accuracy. The probe was inserted in the center of the bore during measurement.

The transport current testing results for various segments of Cu-1 at 4.2 K are displayed in Figure 2. There were 14 layers and voltage taps were placed at layer 2, 4, 6,

8, 10, and 12. Critical current, $I_c$, defined using the 1 µV/cm criterion, is 99 A for the overall coil (C-F). The inner taps reached similar values before quenching partway up their transitions. Figure 3 displays $J_c$, $I_c$, and $B_{coil}$ vs temperature for Cu-1. $J_c$ values of 54.5 kA/cm$^2$ and 31.9 kA/cm$^2$ in self field of the magnet are about a factor of 7 down from the strands with a Nb-chemical barrier (Table 2), as might be expected. Strand and coil $J_e$ values at 4.2 K and 20 K are also given in Table 2.

Figure 4 shows the $J_c$, $I_c$, and $B_{coil}$ vs temperature for NbCu-7A. Strand and coil $J_e$ values at 4.2 K and 20 K are given in Table 2. Here the $J_c$s reach 308 kA/cm$^2$ at 4.2 K and 175 kA/cm$^2$ at 20 K (self field of magnet). Similar analysis is given for NbCu-7B in Figure 5 and Table 2. The $J_c$ values are somewhat lower for this particular strand, however, a slightly larger number of turns on the magnet lead to a higher $B_{coil}$, which reaches 1.57 T at 4.2 K and 0.97 T at 20 K. In order to see how close to the short sample results that the strands in these coils have reached, we will need to first have an accurate calculation of the magnetic field throughout these coils.

**Field Distributions and Load Line for an MgB$_2$ Coil**

Figure 6 shows the schematic of the coil and the winding geometry used to calculate the coil fields. Figure 7 shows the field distribution calculated from numerical methods within coil NbCu-7A. Figure 7 (a) gives a calculation of $B_z$ vs. $z$ at $r = 0$. The calculation gives 1.45 T in the center of the solenoid bore, in reasonable agreement with the 1.5 T measured, and 2.3 T near the innermost winding. Figure 7(b) gives $B_z$ $B_r$, and $B_\theta$ vs. $r$ at the midplane ($z=0$). In order to perform these calculations, we used a numerical method, where the current was assumed uniformly distributed within the strand. For convenience, we chose to work with a strand containing 91 filamentary elements as an approximation to this continuum (Figure 6).

We note that the at $I_c$ = 222 A, the bore field is calculated as 1.45 T (close to the 1.5 T measured) while the calculated maximum field on the windings is about 2.3 T. This enables us to superpose a "calculated" load line for Coil NbCu-7A on the results of an short sample (1 m of strand wrapped helically onto an ITER barrel [25]) measurement of a sample strand from which it was wound, Figure 8.

**Summary**

Three solenoids were wound, one from a monofilamentary $MgB_2$/Cu strand, the other two from seven filament multifilamentary strands with internal Cu-stabilization. Coils NbCu-7A and NbCu-7B were the best performers, reaching short sample performance, and generating 1.5-1.57 T in zero background field at 4.2 K. At 20 K a field of 0.87-0.97 T was generated. $J_c$ values for the strand used in NbCu-7A were 308 kA/cm$^2$ and 175 kA/cm$^2$ at 4.2 K and 20 K respectively. At 20 K the strand $J_e$ for NbCu-7A was 22.7 kA/cm$^2$ and the coil winding pack $J_{e,w}$ was 13.4 kA/cm$^2$. Coil NbCu-7B had overall similar performance, generating 1.57 T at 4.2 K and 0.97 T at 20 K. These coil results show that in-situ route, Cu-stabilized, multifilamentary strands can be made on the 100 m+ length with good performance characteristics and consistent properties.

## Acknowledgements

This work has been funded by an NIH SBIR No. 5710001628 and 1R43EB003752-01, a State of Ohio Technology Action fund, and by DOE HEP grant No. DE-FG02-95ER40900.

## References

[1] G. Grasso, A. Malagoli, M. Modica, et al., Supercond. Sci. Technol. **16,** 271–275 (2003).

[2] G. Grasso, A. Malagoli, D. Marre, et al., Physica C **378–381,** 899–902 (2002).

[3] R Flükiger, P Lezza, C Beneduce, et al., Supercond. Sci. Technol. **16,** 264–270 (2003)

[4] R Flükiger, H.L. Suo, N. Musolino, et al., Physica C **385,** 286–305 (2003).

[5] D. Eyidi, O. Eibl, T. Wenzel, et al., Supercond. Sci. Technol. **16,** 778–788 (2003).

[6] A. Matsumoto, H. Kumakura, H. Kitaguchi, and H. Hatakeyama, Supercond. Sci. Technol. **16,** 926–930 (2003).

[7] Y. Ma, H. Kumakura, A. Matsumoto, et al., Supercond. Sci. Technol. **16,** 852–856 (2003).

[8] B A Glowacki, M Majoros, M Vickers, et al., Supercond. Sci. Technol. **16,** 297–305 (2003).

[9] B.A. Glowacki, M. Majoros, M. Eisterer, et al., Physica C **387,** 153–161 (2003).

[10] E.W. Collings, E. Lee, M.D. Sumption, et al., Rare Metal Mat. Eng. **31,** 406-409 (2002).

[11] E.W. Collings, E. Lee, M.D. Sumption, et al., Physica C **386** 555-559 (2003). See also (for the CTFF process itself) V. Selvamanickam et al., Supercond. Sci. Technol. **8** 587-590 (1995).

[12] S. Soltanian, X.L. Wang, I. Kusevic, et al., Physica C **361**, 84-90 (2001).

[13] M.D. Sumption, M. Bhatia, S.X. Dou, et al. Supercond. Sci. Technol. **17** (2004) 1180-1184.

[14] S. Soltanian, J. Horvat, X.L. Wang, et al., Supercond. Sci. Technol. **16** (2003) L4-L6.

[15] M. Bhatia, M.D. Sumption, M. Tomsic, and E.W. Collings, Physica C **407** (2004) 153-159.

[16] K. Tanaka, M. Okada, H. Kumakura, et al., Physica C **382** (2002) 203-206.

[17] T. Machi, S. Shimura, N. Koshizuka, and M. Murakami, Physica C **392-396** (2003) 1039-1042.

[18] H. Fang, P.T Putman, S. Padmanaghan, et al., Supercon. Sic. Technol. **17** (2004) 717-720.

[19] Y.S. Hascicek, Z. Aslanoglu, L. Arda, et al., Adv. Cryo. Eng. **50** (2004) 541.

[20] A. Serquis, L. Civale, J.Y. Coulter, et al., Supercond. Sci. Technol. **17** (2004) L35-L37.

[21] R. Musenich, P. Fabbricatore and S. Farinon. et. al, "Behavior of $MgB_2$ react and wind coils above 10 K", presentation 4LS01 at the Applied Superconductivity Conference Jacksonville, FL 2004.

[22] M.D. Sumption, M. Bhatia, M. Rindfleisch, J. Phillips, M. Tomsic, and E.W. Collings, to be published in IEEE Trans. Supercond. 2005.

[23] M.D. Sumption, M. Bhatia, M. Rindfleisch, J. Phillips, M. Tomsic, and E.W. Collings, " $MgB_2$/Cu Racetrack Coil Winding, Insulating, and Testing", to be published in Supercond. Sci. Tech. 2005.

[24] M.D. Sumption, M. Bhatia, X. Wu, M. Rindfleisch, M. Tomsic, and E.W. Collings "Multifilamentary, in-situ Route, Cu-stabilized $MgB_2$ Strands:", to be published in Supercond. Sci. Tech. 2005.

[25] L.F. Goodrich and A.N. Srinavasta, in *Critical Currents in Superconductors*, Proc. 7th. Int. Workshop, H.W. Weber ed. (World Scientific Publishing, Singapore, 1994) p. 609.

# List of Tables

Table 1. Strand Parameters.

| Coil Name | Strand Trace ID | SC Fil No. | Strand Construction/Name | Powder add | Strand OD, mm | SC % | Cu % |
|---|---|---|---|---|---|---|---|
| Cu-1 | 566 | 1 | SC/Cu-mono | -- | 1.01 | 23 | 77 |
| NbCu-7A | 573 | 6 | SC/Nb/Cu/monel-6SC-1Cu | Mg[a] | 0.84 | 13 | 32 |
| NbCu-7B | 600 | 6 | SC/Nb/Cu/monel-6SC-1Cu | Mg[a] | 0.84 | 16 | 28 |

[a] Mg was added to form the ratio $Mg_{1.1}B_2$

Table 2. Coil Parameters and Performance Specifications.

| Parameters | Cu-1 | NbCu-7A | NbCu-7B |
|---|---|---|---|
| Strand | SC/Cu mono | SC/NbCu/monel-6SC-1Cu | SC/NbCu/monel-6SC-1Cu |
| Insulation | 1 layer S-glass | 1 layer S-glass | 1 layer S-glass |
| HT (°C/min) | 675/30 | 700/20 | 700/20 |
| SC Fill Factor % | 23 | 13 | 16 |
| Strand/pack fill % | 64 | 59 | 57 |
| Winding pack OD, mm | 104.60 | 104.64 | 108.92 |
| Winding Pack ID, mm | 76.20 | 76.20 | 76.20 |
| Winding Pack Height, mm | 31.75 | 31.75 | 31.75 |
| WP cross sectional area, mm$^2$ | 450.85 | 451.49 | 519.43 |
| No. Turns | 364 | 480 | 538 |
| Total Conductor $L$, m | 104.1 | 137.5 | 155.8 |
| Turns/layer | 26 | 30 | 31/30* |
| No. Layers | 14 | 16 | 17.5 |
| $I_c$, 4.2 K, zero background, A | 99 | 222 | 211 |
| $J_c$, kA/cm$^2$ SC, 4.2 K | 54.5 | 308 | 238 |
| $J_e$ kA/cm$^2$, Strand, 4.2 K | 12.5 | 40.0 | 38.1 |
| $J_{e,w}$ kA/cm$^2$, pack, 4.2 K | 8.1 | 23.6 | 21.9 |
| $B_{coil}$, T, 4.2 K | 0.51 | 1.51 | 1.57 |
| $I_c$, 20 K, zero background, A | 58 | 126 | 126 |
| $J_c$, kA/cm$^2$ SC, 20 K | 31.9 | 175 | 142 |
| $J_e$ kA/cm$^2$, Strand, 20 K | 7.3 | 22.7 | 22.7 |
| $J_{e,w}$ kA/cm$^2$, pack, 20 K | 4.6 | 13.4 | 13.1 |
| $B_{coil}$, T, 20 K | 0.28 | 0.87 | 0.97 |
| % Short Sample (20 K) | -- | 100 | 100 |

* Last layer had 16 turns.

# List of Figures

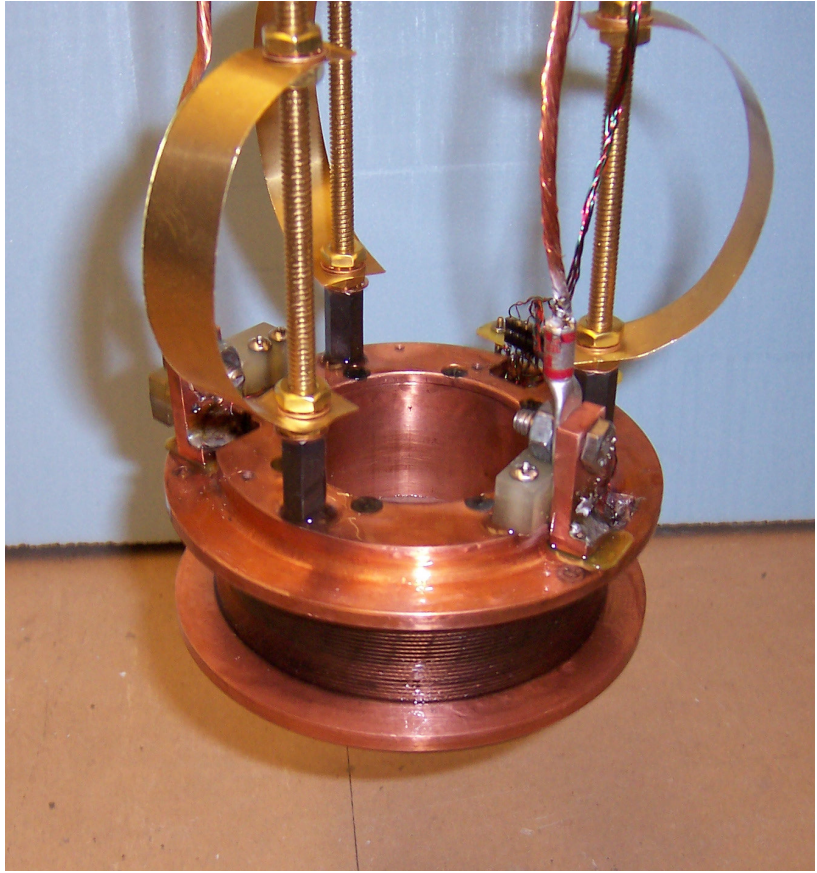

**Figure 1. SUMPTION**

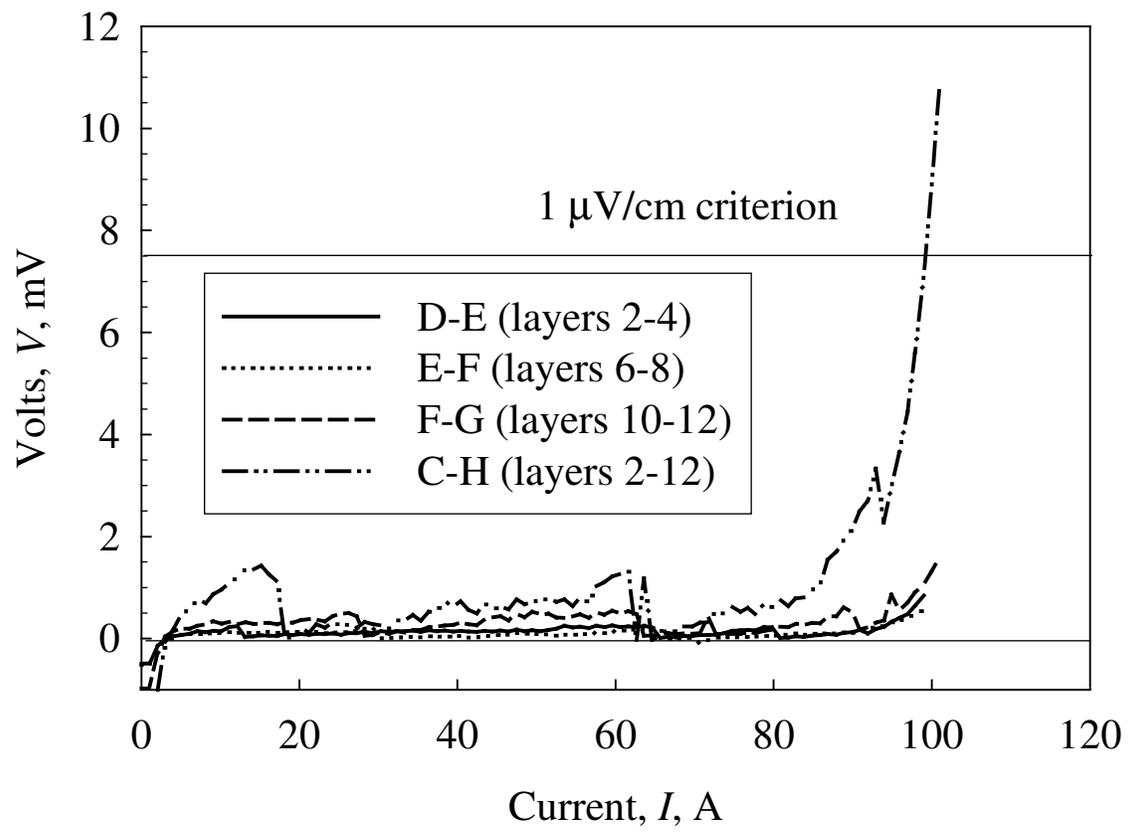

**Figure 2. SUMPTION**

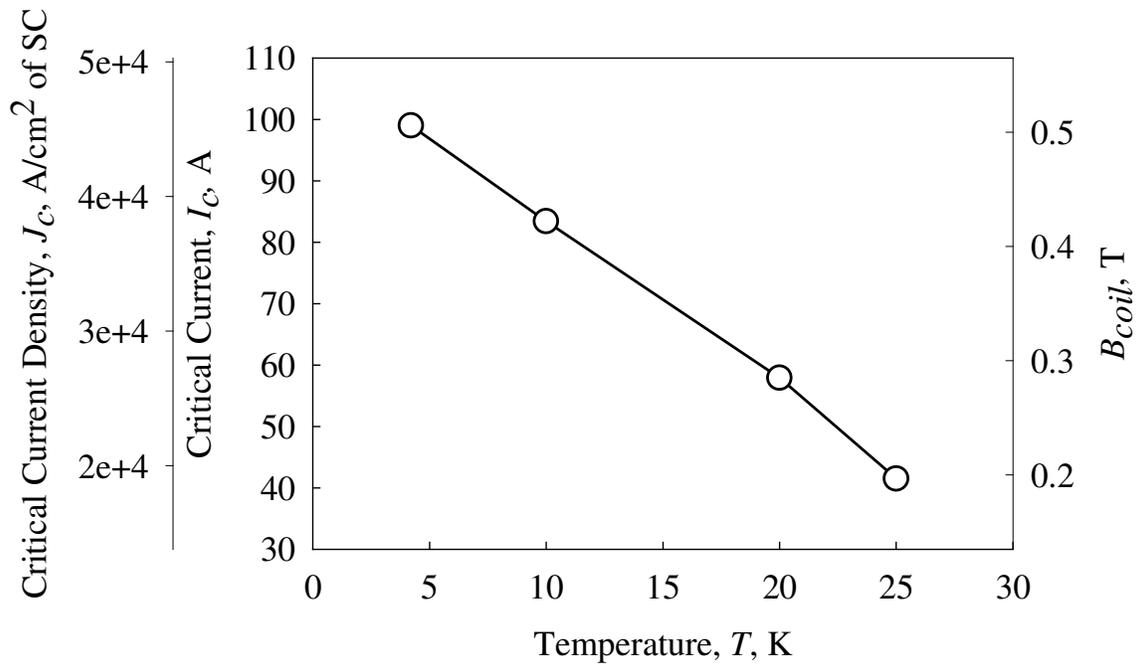

**Figure 3. SUMPTION**

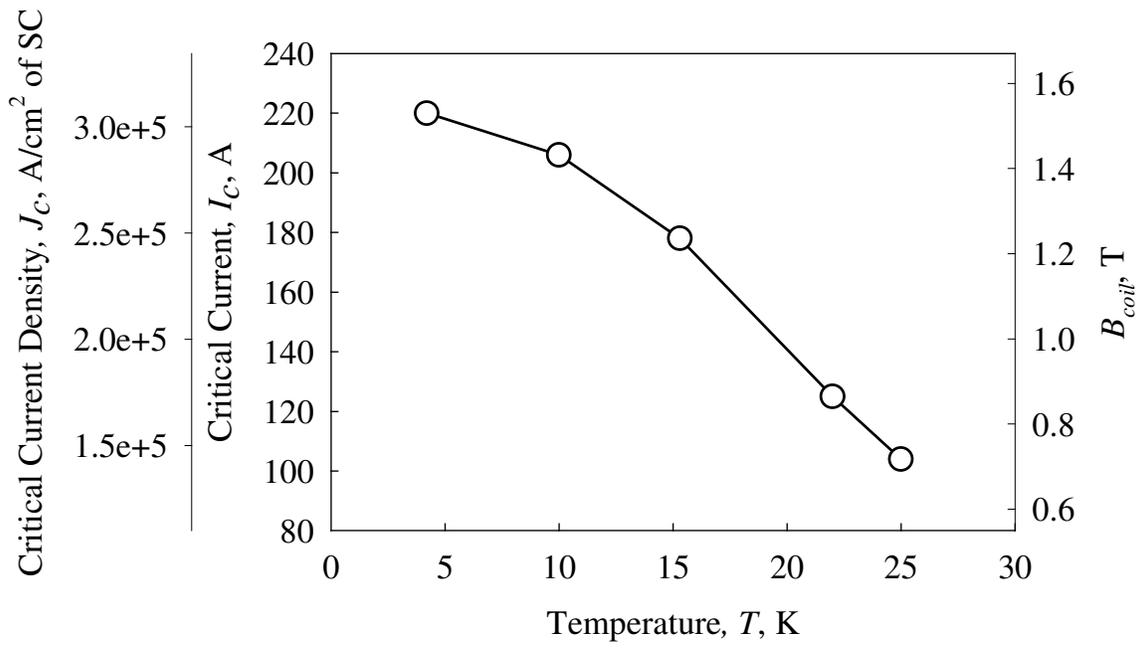

**Figure 4. SUMPTION**

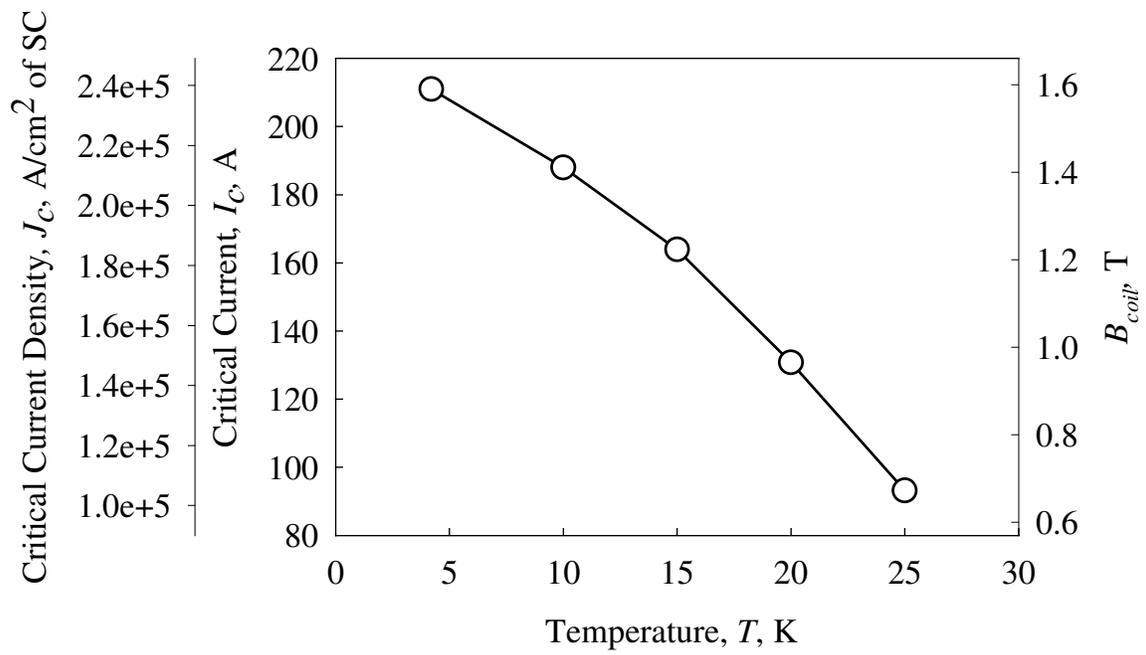

**Figure 5. SUMPTION**

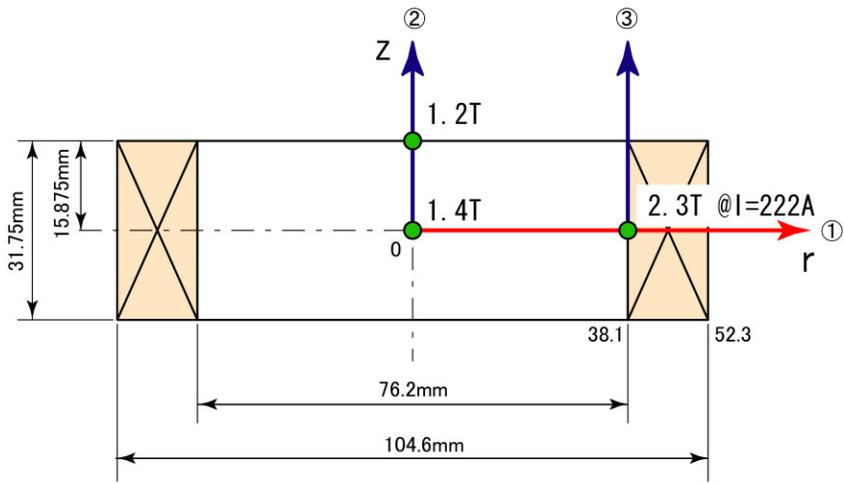

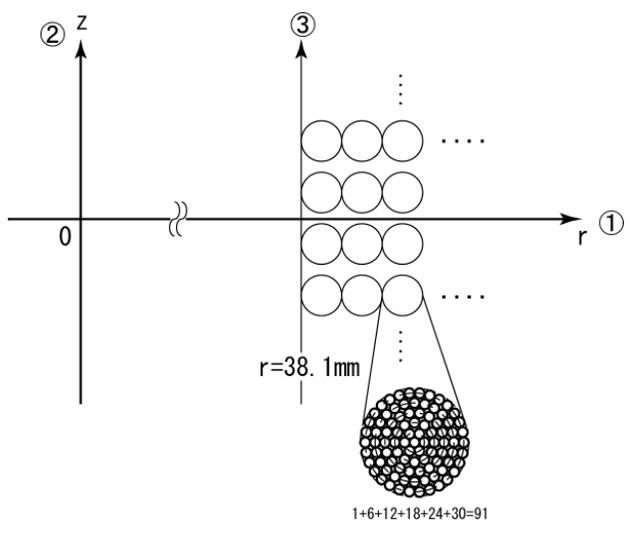

**Figure 6. SUMPTION**

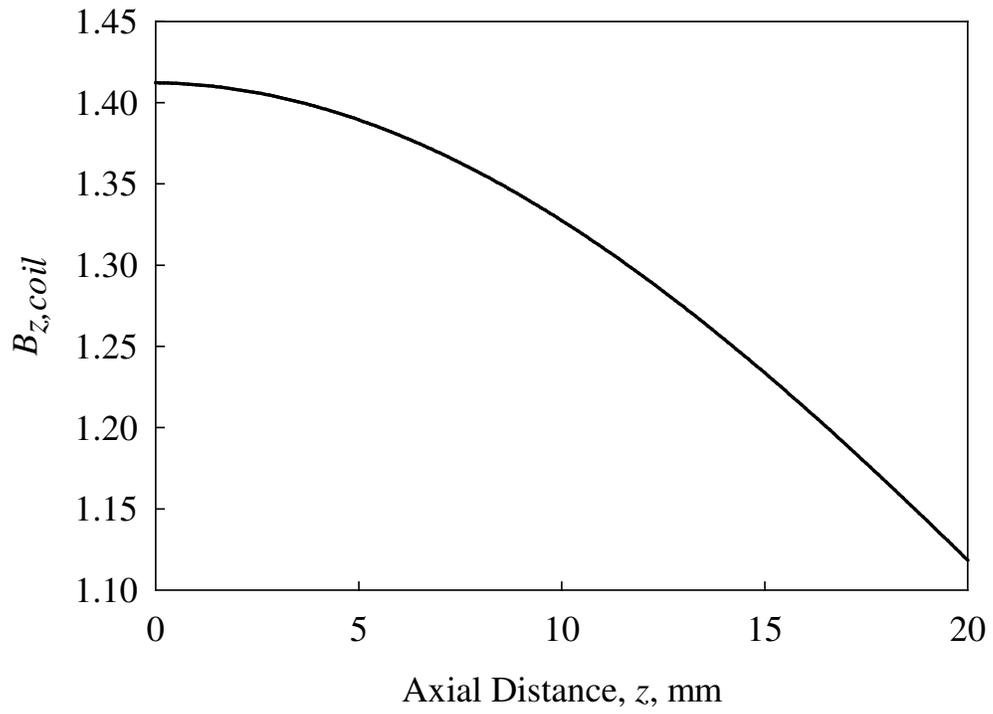

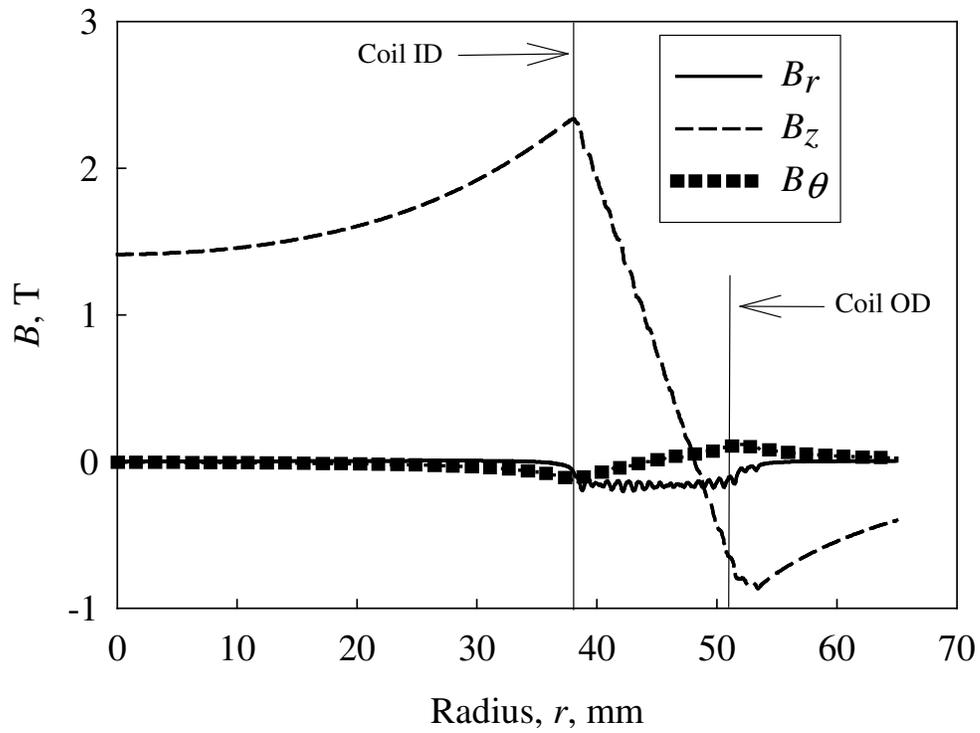

**Figure 7. SUMPTION**

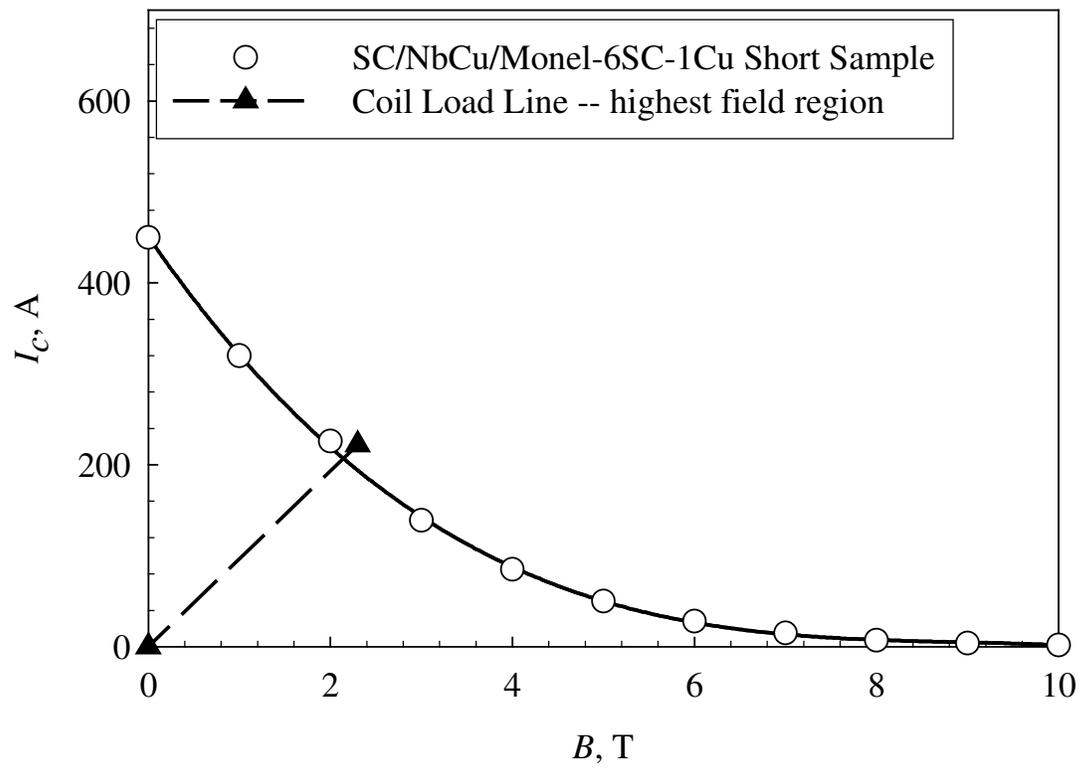

**Figure 8.** SUMPTION